\documentclass[aps,twocolumn,showpacs,nobibnotes,epsf]{revtex4}

\usepackage{graphicx}
\usepackage{dcolumn}
\usepackage{bm}

\begin{document}
\title{Transport properties in Cu$_{x}$TiSe$_{2}$ (0.015$\leq$x$\leq$0.110) Single Crystal}
\author{G. Wu, H. X. Yang, L. Zhao, X. G. Luo, T. Wu, G. Y. Wang }
\author{X. H. Chen}
\altaffiliation{Corresponding author} \email{chenxh@ustc.edu.cn}
\affiliation{Hefei National Laboratory for Physical Science at
Microscale and Department of Physics, University of Science and
Technology of China, Hefei, Anhui 230026, People's Republic of
China\\}

\begin{abstract}

Transport properties are systematically studied for the single
crystals Cu$_{x}$TiSe$_{2}$ (0.015$\leq$x$\leq$0.110). Both of
in-plane and out-of-plane resistivity show the coexistence of
superconducting transition and charge density wave (CDW) for the
single crystals with $x \leq 0.025$. After CDW state is completely
suppressed around x=0.055, the superconductivity is apparently
enhanced by Cu doping. No superconducting transition is observed
above 1.8 K for Cu$_{0.11}$TiSe$_{2}$. Anisotropy in resistivity
increases with increasing Cu content, and is nearly
$T$-independent. CDW state has a strong effect on Hall coefficient
and thermopower. A large thermopower, comparable to the triangle
lattice $NaxCoO_2$, is observed in Cu$_{x}$TiSe$_{2}$.
Intercalation of Cu induces a negative MR due to the interaction
between conducting carries and localized magnetic moments.

\end{abstract}


\vskip 300 pt

\maketitle

\section*{INTRODUCTION}
Layered transition metal dichalcogenides (TMD's) MX$_2$ (M=
transition metal, X=S, Se, or Te) have been extensively studied
due to its two dimensionality structure and physical properties,
such as superconductivity and charge density wave (CDW)
transition.\cite{Wilson,Friend,Yoffe} The structure of these
compounds usually manifests two-types of 1$T$ and 2$H$ phase, both
of structures are formed by X-M-X sandwiches. The top and bottom
sheets are chalcogen atoms, and the middle sheet is metal atom. In
$1T$ phase, all the sandwiches stack to each other as octahedral.
While in $2H$ phase, the sandwiches stack as trigonal
prismatic.\cite{Wilson} CDW state is one of the most interesting
phenomena observed in TMD's and widely studied. Superconductivity
also appears in some materials, such as 2$H$-NbS$_2$ which is a
multi-gap $s$-wave superconductor. Coexistence of
superconductivity and CDW state has been widely observed in 2$H$
transition metal dichalcogenides: $TaSe_2$, $TaS_2$ and $NbSe_2$.
2H-$TaSe_2$ experiences an incommensurate and a commensurate CDW
transitions at 122 K and 90 K, respectively; 2H-$TaS_2$ and
2H-$NbSe_2$ undergo only one incommensurate CDW transition at 75 K
and 35 K; while 2H-$NbS_2$ shows no CDW
transition.\cite{Wilson,Yoffe,moncton} In contrast to the CDW
transition, the superconducting transition temperature decreases
from  2H-$TaSe_2$, through 2H-$TaS_2$ and 2H-$NbSe_2$ to
2H-$NbS_2$, indicating that CDW state and superconductivity
compete to each other.

The interaction between the layers of TMD's is weak van der Waals
type, so that this kind of compounds can be easily intercalated by
various guest species (eg. alkali metal atom, 3$d$ transition
metal atom, or molecules).\cite{Whittingham,Whittinghama,Trichet}
Lattice parameters are usually changed by intercalation,
superstructure can be observed at some guest
concentration.\cite{Yoffe,Masasi} Such superstructure is related
to the CDW state. Recently, it is found that intercalation of Na
into 2$H$-TaS$_2$ increases the superconducting transition
temperature to 5K, and supresses the CDW
transition.\cite{Fang,Sernetz,Lerf} Effect of pressure on
superconductivity and CDW state shows similar result in
2$H$-NbSe$_2$.\cite{Suderow} These results further indicate that
the superconductivity and CDW state compete to each
other.\cite{Yoffe, Castro}

The intercalation of various guests into van der Waals gaps of
MX$_2$ is another way to study TMD's and to find new phenomenon.
Recently, Cu is successfully intercalated into
1T-$TiSe_2$.\cite{Morosan} Intercalation of Cu continuously
suppresses the CDW transition, superconductivity emerges in the
sample $Cu_xTiSe_2$ with $x\sim0.04$, and reaches a maximum
transition temperature ($T_c$) of 4.15 K at x=0.08, then $T_c$
decreases with further doping Cu.\cite{Morosan} Phase diagram of
Cu$_x$TiSe$_2$ is quite similar to that of high $T_c$ cuprates.
Therefore, it is of great interest to reveal if unconventional
superconductivity exists in Cu$_x$TiSe$_2$ system. Cu$_x$TiSe$_2$
is the first superconducting 1$T$ structured MX$_2$ compound.
TiSe$_2$ intercalated with other 3$d$ transition metal (eg: V, Cr,
Mn, Fe, Co, Ni) has been reported, but no superconductivity is
observed.\cite{Titov,Masasi}  Study on anisotropic properties of
Cu$_{0.07}$TiSe$_2$ shows that it was a normal type II
superconductor.\cite{Morosan1} The results of thermal conductivity
for Cu$_{0.06}$TiSe$_2$ indicate that it is a single-gap $s$-wave
superconductor.\cite{Li} ARPES results show that intercalation of
Cu could enhance the density of states, leading to the
superconductivity, but excessive Cu makes superconductivity
disappeared due to strong inelastic scattering.\cite{Zhao} In
order to understand the CDW state and superconductivity in this
novel compound Cu$_x$TiSe$_2$, we made detailed study on transport
properties on high quality single crystals Cu$_x$TiSe$_2$. It is
found the coexistence of superconducting transition and charge
density wave (CDW) for the single crystals with x=0.015 and
x=0.025, different from polycrystalline case in which
superconductivity emerges at x=0.04.\cite{Morosan} The CDW state
shows an apparent effect on resistivity, Hall coefficient and
thermopower.

\section*{EXPERIMENT DETAILS}

  High quality Cu$_x$TiSe$_2$ single crystals were grown by the
chemical iodine-vapor transport method. Powders of Cu (99.7\%), Ti
(99.5\%) and Se (99.5\%) were mixed and thoroughly ground, then
pressed into pellet. The pellet was sealed under vacuum in a
quartz tube with diameter of 13mm and length of 150mm with iodine
(10mg/cm$^{-3}$). One end of tube was slowly heated to 940$^o$C
with heating rate of $\sim$150$^o$C/h, another end of the tube was
kept at 740$^o$C. After one week, the furnace was cooled to room
temperature over a few hours. Finally many golden plate-like
Cu$_x$TiSe$_2$ crystals were obtained. The typical dimensional is
about 5$\times$5$\times$0.03mm$^3$. Because the element Cu can
react with Se very easily, excessive Cu was used as starting
material. For example, in order to grow the Cu$_{0.055}$TiSe$_2$
crystals, the stoichiometry of Cu, Ti and Se powder is  0.4:1:2 in
molar ratio. Actual composition of the single crystals was
determined by X-ray Fluorescence Spectroscopy (XRF) (XRF-1800,
Shimadzu Inc.)

All crystals were characterized by Rigaku D/max-A X-Ray
diffractometer (XRD) with graphite monochromatized Cu K$\alpha$
radiation ($\lambda$=1.5406\AA) in the 2$\theta$ range of
10$-$70$^{\circ}$ with the step of 0.02$^{\circ}$ at room
temperature. Resistivity measurements were performed on a AC
resistance bridge (Linear Research, Inc., Model LR700) by the
standard four-probe method. The magnetic field was supplied by a
superconducting magnet system (Oxford Instruments). Hall contact
configuration is the standard ac six-probe geometry. To eliminate
the offset voltage due to the asymmetric Hall terminals, the
magnetic field was scanned from -5 to 5 T. Thermopower
measurements were carried out with a home-built apparatus and
performed using small and reversible temperature differences of
0.5K. Two ends of the single crystals were attached to two
separated copper heat sinks to generate the temperature gradient
along the crystal ab plane. Two Rh-Fe thermometers were glued to
the heat sink (next to the single crystals). Copper leads were
adhered to the single crystals and all the data were corrected for
the contribution of the Cu leads.

\section*{RESULTS AND DISCUSSION}

\begin{figure}[h]
\centering
\includegraphics[width=9cm]{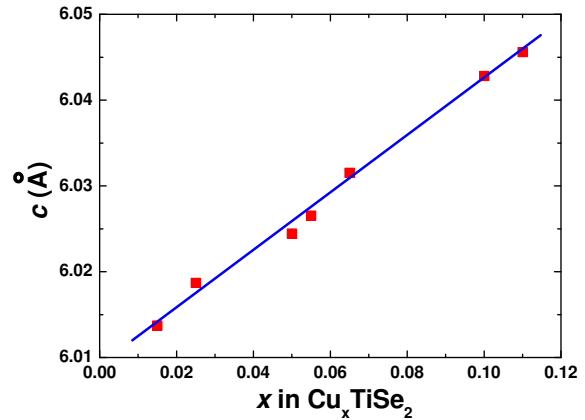}
\caption{C-axis lattice parameter as a function of $x$ in
Cu$_{x}$TiSe$_{2}$ for single crystals, being consistent with that
reported in polycrystalline samples.\cite{Morosan} The straight
line guides eyes.} \label{fig1}
\end{figure}

Figure 1 shows c-axis lattice parameter as a function of Cu
content (x). The c-axis lattice parameter was obtained by the
X-ray diffraction on single crystals Cu$_xTiSe_2$ with different
$x$. It shows a good relationship between the c-axis lattice
parameter and Cu content. The c-axis lattice parameter increases
linearly with increasing Cu content. This is consistent with that
reported in polycrystalline samples.\cite{Morosan}

\begin{figure}[b]
\centering
\includegraphics[width=9cm]{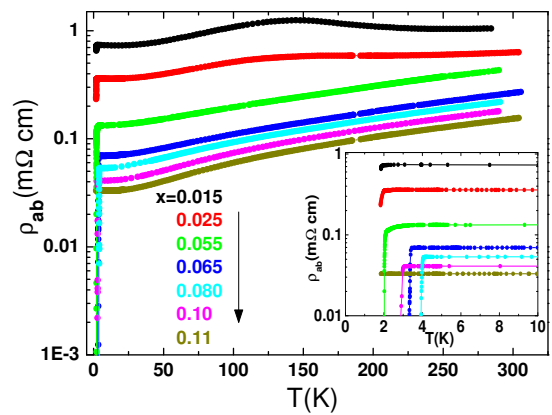}
\caption{Temperature dependence of in-plane resistivity down to
1.8 K for Cu$_{x}$TiSe$_{2}$ crystals with different Cu contents.
The low temperature $\rho(T)$ data are plotted in the inset.}
\label{fig2}
\end{figure}

Figure 2 shows temperature dependence of in-plane resistivity
$\rho_{ab}$ from 300 K to 1.8 K for Cu${_x}$TiSe$_2$ crystal with
different Cu contents (x=0.015, 0.025, 0.055, 0.065, 0.08, 0.10
and 0.11). As show in Fig.2(a), $\rho$$_{ab}$ shows a CDW hump for
the samples with $x\leq0.025$, the CDW hump becomes broader and
moves to lower temperature with increasing $x$. For the sample
with x=0.055, the CDW hump disappears and the resistivity shows
nearly $T^2$-dependent. It suggests that CDW state is completely
suppressed around x=0.055. Intercalation of Cu leads to
suppression of CDW state could be related to the structural change
induced by Cu doping. In TiSe$_2$, the TiSe$_2$ goes through a
2$\times$2$\times$2 structural transition below 200 K, leading to
a CWD state.\cite{Kidd,Salvo} The CDW state has a close
relationship with the structure of TiSe$_2$. When Cu is
intercalated into the TiSe$_2$ layers, such 2$\times$2$\times$2
lattice is destroyed. The disorder of the Cu atoms destroys the
superlattice of TiSe$_2$, leading to suppression of the CDW state.

\begin{figure}[h]
\centering
\includegraphics[width=9cm]{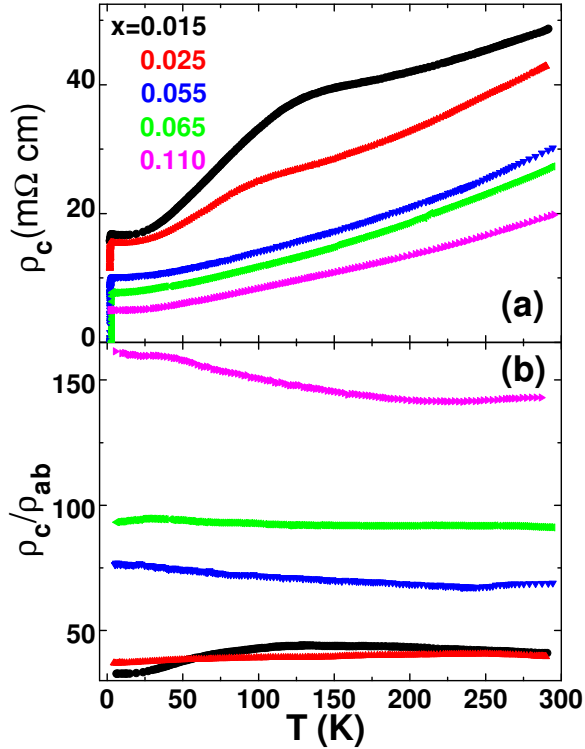}
\caption{(a): Temperature dependence of out-of-plane resistivity
for Cu$_{x}$TiSe$_{2}$ crystals with different Cu contents;
(b):Temperature dependence of anisotropy in resistivity for
Cu$_{x}$TiSe$_{2}$ crystals with different Cu contents.}
\label{fig3}
\end{figure}

The resistivity monotonically decreases with increasing Cu
content. The residual resistivity ratio
$\rho$$_{ab}$(300K)/$\rho$$_{ab}(5K)$ increases from $\sim 2$ to
$\sim 5$ with increasing Cu content from 0.015 to 0.11. These
results are consistent with that reported in polycrystalline
sample.\cite{Morosan} In contrast to the results of
polycrystalline samples, the samples with CDW state shows
superconducting transition although the resistivity does not reach
zero down to 1.8 K. In order to clearly show the superconducting
transition, the low temperature $\rho_{ab}(T)$ data were plotted
in the inset of Fig.2. The superconducting transition temperature
($T_{onset}$) is 1.92, 2.03, 2.24, 3.40, 4.13 and 3.02 K for the
samples with x=0.015, 0.025, 0.055, 0.065, 0.08 and 0.10,
respectively. It indicates that the superconductivity and CDW
state coexist in the samples with slight Cu doping ($x\leq
0.025$), which is similar to the case of $2H$
family.\cite{Wilson,Yoffe,moncton} In CDW regime, $T_{onset}$
slightly increases with increasing Cu content, while CDW
temperature decreases. It implies a competition of
superconductivity and CDW state, similar to the case of 2H family.
It should be pointed out that superconducting transition is not
observed in the sample with x=0.11 with cooling the sample to 1.8
K. These results are nearly consistent with the phase diagram
obtained from polycrystalline samples\cite{Morosan} except for the
observation of superconducting transition for the slightly
Cu-doped samples with CDW state.

Temperature dependence of out-of-plane resistivity $\rho_{c}$ is
plotted in Fig.3(a) for the samples Cu${_x}$TiSe$_2$ crystal with
different Cu contents (x=0.015, 0.025, 0.055, 0.065 and 0.110). As
shown in Fig.3(a), $\rho_{c}(T)$ shows similar behavior to
$\rho_{ab}(T)$. $\rho_{c}(T)$ monotonically decreases with
increasing Cu content. An obvious CDW hump can be observed in
$\rho_{c}(T)$ for the slightly Cu-doped samples.  The CDW state is
suppressed by intercalation of more Cu. $\rho_{c}(T)$ shows
$T^2$-dependent behavior for the samples without CDW state.
Anisotropic $\rho$$_{ab}$/$\rho$$_{c}$ are shown in Fig.3(b) for
the samples Cu${_x}$TiSe$_2$ crystal with different Cu contents
(x=0.015, 0.025, 0.055, 0.065 and 0.110).
$\rho$$_{ab}$/$\rho$$_{c}$ shows a weak temperature dependence in
all the samples with different Cu contents. It indicates that the
charge transport mechanism is the same in $ab$-plane and along
Cu${_x}$TiSe$_2$. Compared to the case of $TiS_2$, the anisotropy
for the samples Cu${_x}$TiSe$_2$ is smaller. In the case of
$TiS_2$£¬ the resistivity-anisotropy is 1500 and 750 at 5 K and
300K, respectively.\cite{Imai} In Cu${_x}$TiSe$_2$, Se has larger
radius and stronger covalence than S,  so that the carriers can
move more easily between the layers in Cu${_x}$TiSe$_2$ than
TiS$_2$. With intercalation of Cu, in-plane conductivity is
strongly enhanced than $c$-axis conductivity, so that the
anisotropy increases with increasing $x$ in Cu${_x}$TiSe$_2$. A
similar enhancement of anisotropy with increasing carrier content
is observed in n-type cuprates.\cite{wang}
\begin{figure}[b]
\centering
\includegraphics[width=9cm]{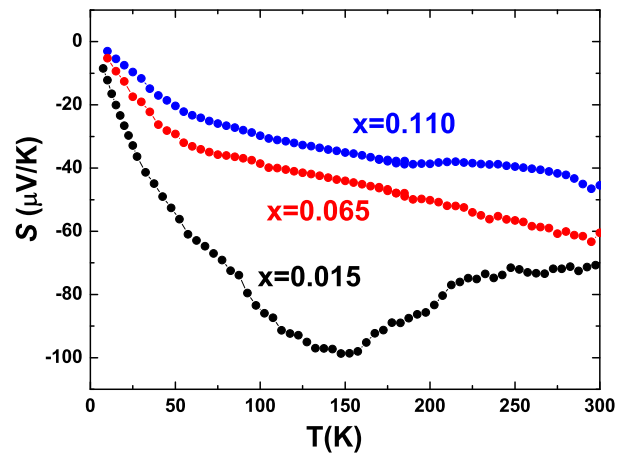}
\caption{Temperature dependence of in-plane thermopower for the
samples Cu$_{x}$TiSe$_{2}$ with $x$=0.015, 0.065, 0.110.}
\label{fig4}
\end{figure}

Figure 4 shows temperature dependence of in-plane thermopower
($S$) for the  Cu${_x}$TiSe$_2$ crystals with $x$=0.015, 0.065 and
0.110. Thermopower is negative in the whole temperature range. It
indicates that the carrier is n-type. At room temperature, the
magnitude of $S$ decreases with increasing x. Thermopower shows a
non-monotonic temperature dependence for the sample
Cu$_{0.015}$TiSe$_{2}$, a maximum value of $S$ appears at 150 K at
which a CDW transition is observed in resistivity. It suggests
that the non-monotonic T-dependence for $S$ arises from the
occurrence of CDW order. For the samples without CDW state, the
magnitude of $S$ monotonically decreases with decreasing
temperature. All these results indicate that the metallic nature
of Cu$_{x}$TiSe$_{2}$ is enhanced by the electron doping. It
should be pointed out that the thermopower is anomalous large.
Compared to the triangle lattice $NaxCoO_2$ with large
thermopower, Cu${_x}$TiSe$_2$ shows a larger $S^2/\rho_{ab}$ at
room temperature than $NaxCoO_2$. In Cu$_{0.065}$TiSe$_2$, $S(300
K)\simeq60 \mu V/K$ and $\rho_{ab}(300 K)\simeq0.15 m\Omega cm$;
while for $Na_{0.68}CoO_2$, $S(300 K)\simeq 90 \mu V/K$ and
$\rho_{ab}(300 K)\simeq 1 m\Omega cm$.\cite{Terasaki} In this
sense for thermoelectirc material, Cu${_x}$TiSe$_2$ have more
larger anomalous thermopower than $NaxCoO_2$. It deserves further
investigation.

\begin{figure}[h]
\centering
\includegraphics[width=9cm]{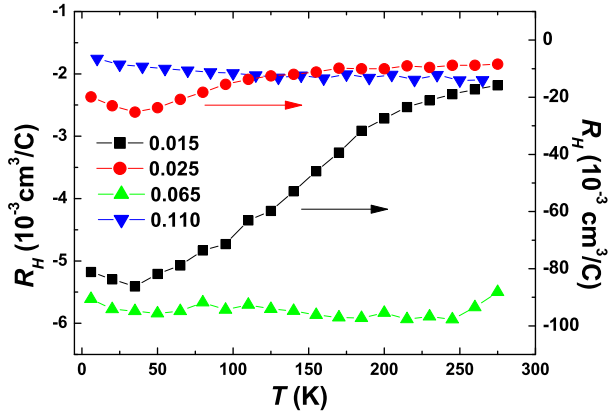}
\caption{Temperature dependence of the Hall coefficient ($R_H$)
for the samples Cu$_{x}$TiSe$_{2}$ with $x$=0.015, 0.025, 0.065,
0.110.} \label{fig5}
\end{figure}

Figure 5 shows temperature dependence of the Hall coefficient
($R$$_H$) for Cu$_{x}$TiSe$_{2}$ with $x$=0.015, 0.025, 0.065,
0.110. In the whole temperature range, the $R_H$ are negative.
Both of thermopower and Hall coefficient indicate that the carrier
is n-type in Cu$_{x}$TiSe$_{2}$ system. The Hall coefficient
decreases with increasing Cu content. It indicates that the
carrier concentration increases with Cu doping, being consistent
with the results of resistivity and thermopower. The samples with
CDW state shows a strong temperature dependent $R_H$ behavior. The
Hall coefficient shows a strong T-dependence below 200 K for the
sample with x=0.015, while below 100 K for the sample with
x=0.025. These results coincide with occurrence of the CDW order
observed in resistivity. For the samples without CDW state, the
Hall coefficient shows T-independent behavior within uncertainty.
It indicates characteristic of metal for the heavily Cu-doped
sample. It further indicates that the T-dependent Hall coefficient
arises from the CDW state. Formation of the CDW state leads to
localization of some carriers, so that the conducting carriers
decreases, and consequently enhances the Hall coefficient. This is
why the $R_H$ increases below 200 K and 100 K for the samples with
x=0.015 and 0.025 as shown in Fig.5.

Figure 6 shows $H^2$-dependence of isothermal in-plane
magnetoresistance (MR) for Cu$_{x}$TiSe$_{2}$ with $x$=0.015,
0.055, 0.065 and 0.110 at 6 K, 15 K and 30 K.  For the sample with
$x$=0.015 , the isothermal MR is positive, and shows a good
$H^2$-dependent behavior at different temperatures. It implies
that the isothermal MR just comes from the contribution of the
Lorentz force on the carriers under magnetic field. However, the
isothermal MR shows a complicated magnetic field (H) dependence
\begin{figure}[b]
\centering
\includegraphics[width=9cm]{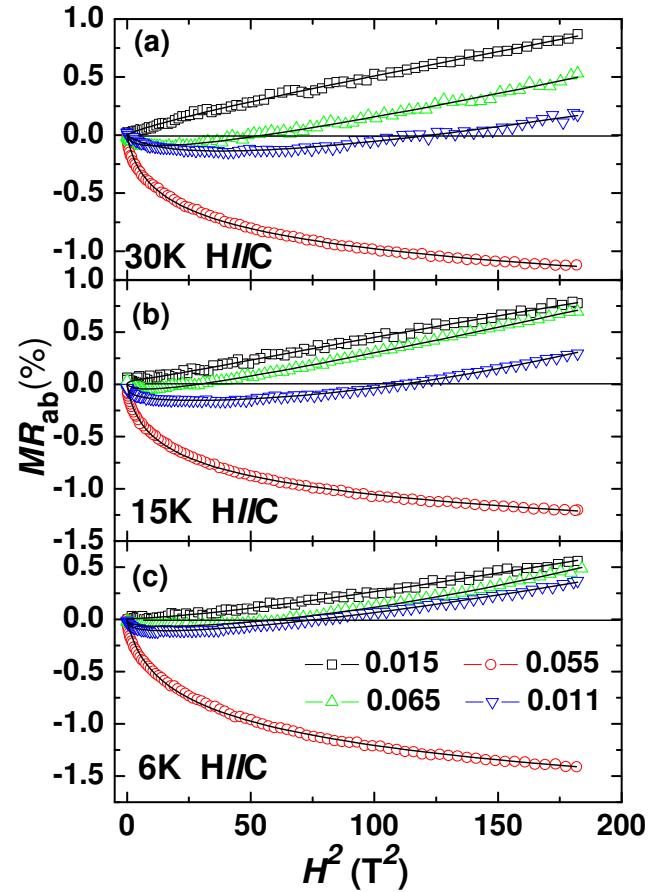}
\caption{Isothermal magnetoresistance  at 6 K, 15 K and 30 K for
the samples Cu$_{x}$TiSe$_{2}$ with $x$=0.015, 0.055, 0.065,
0.110.} \label{fig6}
\end{figure}
for the samples with $x$=0.065 and 0.110. The MR is negative and
increases with increasing H at low magnetic field, then decreases
and crosses zero at certain H, and positive MR increases with
increasing H at high magnetic field. It indicates that the MR
comes from two contributions: one negative and one positive
component. As shown in Fig.6, the isothermal MR shows almost
$H^2$-dependence at high H for  the samples with $x$=0.065 and
0.110. It suggests that the MR is dominant by Lorentz force. In
order to understand the complicated MR, the formula $\Delta
\rho/\rho=-A_1^2ln(1+A_2^2H^2)+B_1^2H^n$ is used to fit the
experimental data. The first term in the formula comes from a
semiempirical expression proposed by Khosla and
Fischer\cite{Khosla} and has been used to explain the negative MR.
The basis for this formula is Toyazawa's localized-magnetic-moment
model of magnetoresistance, where carriers in an impurity band are
scattered by the localized spin of impurity atoms.\cite{Toyazawa}
It is found that all data can be well fitted by the formula for
all samples. It indicates that the negative comes from the
interaction between conducting carriers and localized magnetic
moments. It reveals a decrease of spin-dependent scattering of
carriers in magnetic fields. When Cu is intercalated into the
TiSe$_2$ layers, some of non-magnetic Ti(IV) change into magnetic
Ti(III), so that the interaction between conducting carries and
localized magnetic moments takes place. Such interaction between
conducting carriers and localized magnetic moments results in the
negative contribution to MR. Therefore, the complicated
$H$-dependent MR can be well understood. However, an anomalous MR
is observed for the sample with x=0.055 compared to the
observation of MR in other samples. The sample with x=0.055 shows
negative MR in the whole H range, and the negative MR
monotonically increases with increasing H. It indicates that the
MR is dominant by the interaction between conducting carries and
localized magnetic moments even at high magnetic field. Such
anomaly could be related to the critical point for disappearance
of CDW state.

\section*{CONCLUSION}
 In conclusion, we have grown high quality of
Cu${_x}$TiSe$_2$ single crystals with different Cu contents from
$x$=0.015 to $x$=0.110. The transport properties, anisotropic
resistivity, thermoelectric power, Hall coefficient and
magnetoresistivity are systematically studied. A systematic
evolution of CDW state and superconducting state with Cu content
is observed. It is found that the CDW state can be suppressed by
the intercalation of Cu, while superconductivity is induced.
Before the CDW state is completely suppressed, the CDW state and
superconductivity can coexist and compete each other. The CDW
state gives a strong effect on the transport properties, leading
to T-dependent Hall coefficient. A anomalous large thermopower is
observed, even comparable to large thermopower material
$NaxCoO_2$. Intercalation of Cu induces a negative MR due to the
interaction between conducting carries and localized magnetic
moments.

\section*{ACKNOWLEDGMENTS}
This work is supported by the grant from the Nature Science
Foundation of China and by the Ministry of Science and Technology
of China (973 project No: 2006CB601001) and by National Basic
Research Program of China (2006CB922005).

\end{document}